\documentstyle[aps]{revtex}
\begin{document}
\twocolumn
\title{ Quantum state estimation }
\author{
Z. Hradil \cite{optika} }
\address{ Atominstitut der \"{O}sterreichischen  Universit\"{a}ten,
Sch\"{u}ttelstrasse 115,
A--1020 Wien, Austria
}
\date{\today}
\maketitle

\begin{abstract}
New algorithm for quantum state estimation   based on
 the maximum  likelihood estimation  is proposed.
Existing techniques for state reconstruction
based on the  inversion of measured  data are shown to be
overestimated since they   do not  guarantee the positive
definiteness of the reconstructed density matrix.

\end{abstract}


State reconstruction   belongs to
 the topical problems of contemporary quantum theory.
This sophisticated technique is trying to  determine the  maximum
amount of information about the system--its quantum state.
Even if the history of the problem may be traced back to the early
days of quantum mechanics,  till quantum optics opened
the new era of the  state reconstruction. Theoretical prediction
of Vogel and Risken  \cite{VR} was  closely followed  by the  experimental
realization of the suggested algorithm by Smithey et. al. \cite{SBRF93}.
Since that time  many improvements and new
 techniques have been proposed
 \cite{LP94,AMP94,ALP95,RW95,KWV95,P96,WV96,BW96,LR96,PTKJ96},
to  cite without
 requirements to completeness at least some titles from the existing
 literature \cite{buzek}.  Even if the method
comes  from    optics, similar methods are  currently
being used also in atomic physics  as quantum endoscopy \cite{endo}.
Homodyne detection of quadrature operator with varying phase of
local oscillator  $(x_{\phi},\phi)$ was used as the measurement
in the original  proposal  \cite{VR,SBRF93}.
 The algorithm served  for
 determination of the  Wigner function $ W(x,p)$ and also other
 quasiprobabilities representing the density matrix.
 Measurement of rotated quadrature operator may also be
used for direct evaluation of the coefficient of density matrix in
number state  representation $\rho_{m,n}$ \cite{AMP94} and for the
analysis of  multimode  fields \cite{KWV95}.
Simultaneous  measurement of the
pair of quadrature operators  $(x,p)$ using double homodyne or heterodyne
detection  yields directly the  $Q$--function
  $Q(\alpha)$ \cite{P96}. Surprisingly easy technique was  suggested by
Walentovitz  and Vogel
\cite{WV96} and  by Banaszek and  Wodkiewicz  \cite{BW96}.
Mixing of the signal and coherent fields with
controlled  amplitude   on the  beam splitter may serve for  reconstruction
of the Wigner function and other distributon functions
using the photon counting only.
Techniques similar to the  quantum state reconstruction have been
suggested for indirect observations of particle number, see for example
Ref. \cite{indirect}.
 Though the techniques are different as far as the practical
  realization is concerned, they  all
may be comfortably  represented by  the formalism of generalized
measurement \cite{Hel76}.
As well  known,  any  measurement
 may be described  using the probability operator measure  (POM)
$ \hat \Pi(\xi),$  being  any
 positively defined resolution of identity operator
$ \hat \Pi (\xi) \ge 0,$ $ \int d\xi \hat  \Pi(\xi)  =\hat 1.$
Probability distribution of the outcome predicted  by quantum  theory
is
\begin{equation}
w_{\rho}(\xi) = {\rm Tr}[\hat \rho \hat \Pi(\xi)],
\label{mereni}
\end{equation}
where $\hat \rho$ is the density matrix of the state.
The measured  variable $\xi$ represents  formally the
registered data being in general
a multidimensional vector with the   components
belonging to  both the discrete  and continuous  spectrum as shown in
above mentioned examples. The  key point   of the  existing
reconstruction techniques--inversion  of the relation (\ref{mereni}),
  represents  nontrivial problem. Solution  may be  formally   written
as an analytical identity
\begin{equation}
W_{\rho}(\alpha) = \int d \xi K(\alpha,\xi) w_{\rho}(\xi),
\label{inversion}
\end{equation}
  $W_{\rho}(\alpha)$ being a representation of density matrix.
In order to find the representation $W(\alpha)$
 of a  density matrix corresponding to an unknown signal,
 the existing reconstruction techniques apply  the relation
(\ref{inversion})  on the
actually detected statistics $w(\xi). $

Apart from the fact how ingeniously the  individual inversions have been
done, this treatment is   essentially improper for the  application in
quantum theory.    Particularly, it may represent
 a density matrix only for  such
  measurable   probability distributions, which are  given   {\em exactly}
by the relation  (\ref{mereni}).
Deviations between actually detected  $w(\xi)$  and
the true  statistics $w_{\rho}(\xi)$ are not allowed,
 since they may  spoil the positivity  of reconstructed
density matrix.
This algorithm anticipates therefore the  absolute  precision
impossible in  quantum theory. There  are at least the following
 imperfections of detected  statistics  $w(\xi)$,
 which should be taken into account:
 i)  Sampling error caused by the limited number of
 available  scanned  positions of   continuous variable at which the
 measurement was done.  ii) Counting error caused by   limited set of
 available data counted at each position.
 For example, in  Ref. \cite{SBRF93} the former one is
 caused   by  the   division  of  quadrature
$x_{\phi}$  into  $64$ bins and   phase into $27$ values,
 whereas the later one  by  the detection  of  quadrature   $x_{\phi}$
at each bin.
Other errors such as imperfections of detectors or external noises,
   may appear  in practise as well.
 In quantum  case   the algorithm based on  inversion
 provides a result, but
 does not guarantee the positive definiteness of the reconstructed
 density matrix  \cite{note}.
   In the example of Ref.  \cite{SBRF93}
 the positive definiteness  of reconstructed  matrix
 has not  been checked explicitly, but  can be
judged according to  the  papers  \cite{AMP94,ALP95,LM96}. Here  the
negative part of  photocount  distribution indicates  the spoiling of
  positive definiteness.  Even if there is  a
connection between  the dimension of Hilbert space where this happens
and the  number of phases  \cite{LM96}, a  righorous  way how to treat
the positive definiteness within reconstruction has not yet been
suggested.  As pointed out   by Jones \cite {Jones} in his
Ref. 12, the failure of similar methods  is the rule rather
than the exception    in the case    where the data
 underdetermine the solution.
This happens also  in the case of more dense data, even if
 the  region of ill behaviour is   shifted to less obvious
manifestations.   These methods are
considered as satisfactory only because some additional  information
in the form of data smoothing is used and mathematical difficulties are
neglected.
 Instead of inversion of the detected data, a technique
 motivated by  quantum information theory \cite{Hel76,CD94} and
 by  phase shift
estimation  \cite{phase},  will be suggested in this Rapid
Communication. Previous reconstruction techniques will be embedded
into the common  scheme  based on maximum likelihood estimation.

Many parameters characterizing the
quantum state should be estimated in state reconstruction. As pointed out
by Helstrom \cite{Hel76} this may be done  restricting  the dimension of
Hilbert  space, and accepting some residual
uncertainty.  Similarly,   Jones \cite{Jones} investigated
the fundamental limitations of quantum state measurement using Bayesian
methodology. On the contrary, the
realistic   measurements as in the existing techniques will be
anticipated  here.
Assuming the repeated (or multiple) measurement performed on the   $n $
copies of the system, the output of  observation may be parametrized
by  the set of states (projectors)
  formally denoted as $ |y_1\rangle, \ldots, |y_n\rangle, $
repetition of  a particular outcome  being allowed.
Pure states represent  here
 the case of sharp measurement, whereas unsharp measurement
  involving the
 finite resolution should be  represented   by an appropriate POM.
 Since formal considerations are valid for both
 these cases, the  notation of sharp measurement will
be kept in the following  for the sake of simplicity.
 Maximum likelihood estimation ascribes to  such a
measurement the state  $\hat \rho $ maximizing the likelihood functional
\begin{equation}
{\cal L}(\hat \rho) =  \prod_i^n \langle y_i |\hat \rho |y_i \rangle.
\label{likeli}
\end{equation}
The aim of  this contribution  is to find this state and to clarify
the  fluctuations of such a prediction.
As  the mathematical tool, the inequality between the geometric and
arithmetic averages
  of non--negative numbers $q_i$  will be used
$
\bigl( \prod_i^n\nolimits q_i  \bigr)^{1/n}   \le   \frac{1}{n}
\sum_i^n\nolimits {q_i} .
$
The equality is achieved if and  only if  all  the numbers $q_i$ are
equal.  The variables will be formally  replaced
by $q_i = x_i/a_i,$   where
$x_i \ge 0$ are  positive  and $ a_i>0 $ are auxiliary positive  nonzero
numbers.
In the following the  $n$ dimensial vectors will be denoted by boldface
as  $\bf a$, $\bf x$, $\bf y$, etc..
Assume now that the numbers $q_i$ are chosen from the given set of
values so that the value $q_i$ appears $k_i$ times in the collection of
$ n $ data. Hence $k_i$ represents the frequency,
 $ f_i = k_i/n $ being the  relative frequency
 $ \sum_i\nolimits^{\prime} f_i = 1.$
Parametrization revealing explicitly the  frequency
will be denoted  by  upper  prime in sums and products, indicating
 that index runs over
spectrum of different values.
Without loss of generality the variable $\bf x$ may be interpreted as
probability
$ \sum\nolimits^{\prime}_i x_i = 1, $    since the normalization may
always be  involved in auxiliary variables
 ${\bf a}.$
The  relation,  known as  Jensen's inequality
 \cite{Fuchs},  then reads
\begin{equation}
\prod_{i}\nolimits^{\prime}  \biggl[\frac {x_i}{a_i}  \biggr]^{f_i}
\le    \sum\nolimits_{i}^{\prime}  f_i
\frac{x_i}{a_i}.
\label{qineq}
\end{equation}
  In this form it represents remarkably  powerfull relation
 since  the equality sign may be achieved  for  an arbitrary
 probability ${\bf x  = \bf a  }.$
  For example,  the Gibbs  inequality   \cite{CD94}
  follows  as a special case  chosing the parameters
$
 a_i = f_i,
$
 since the inequality (\ref{qineq}) may be rewritten as
$
- \sum\nolimits_{i}^{\prime} f_i \ln \frac  {f_i} {x_i}
\le   0.
$
These formal manipulations  are tightly connected
to the  maximization of likelihood function.
Using the  definition
  \begin{equation}
    x_i =  {  \langle y_i |\hat \rho |y_i\rangle},
\label{definition}
\end{equation}
 $a_i$  being a  subject of further  considerations, the likelihood
 functional may  be simply estimated as
\begin{eqnarray}
\bigl( {\cal L}(\hat \rho )  \bigr )^{1/n} =  \prod_i\nolimits^{\prime}
\biggl( \langle y_{i} |\hat \rho |y_{i} \rangle \biggr)^{  f_{i}}
   \le      \prod_j\nolimits^{\prime}  a_j^{ f_j}
  {\rm Tr}  \{ \hat \rho  \hat R ({\bf y}, {\bf a})\}.
 \label{true1}
\end{eqnarray}
The operator $\hat R$ is  given, in general  by nonorthogonal,
decomposition  as
\begin{equation}
 \hat R ({\bf y}, {\bf a})  = \sum_i\nolimits^{\prime} \frac{f_i}{ a_i}
 |y_i\rangle \langle y_i|.
 \label{R}
 \end{equation}
 Relation (\ref{true1})  simply follows from the definition (\ref{likeli})
 and from  the inequality (\ref{qineq}).
Further  treatment is distinguished  by the  following  specification of
  auxiliary parameters $\bf  a$:

{\em Reconstructions of wave function}

\noindent
Condition  $a_i = f_i$ tends to considerable simplifications.
Since the measurement need not be complete $  \hat R ({\bf y}, {\bf a }
={\bf f}) \le \hat 1,$  the
right-hand side  of the relation (\ref{true1}) reads
\begin{equation}
(\ref{true1}) = \prod_j\nolimits^{\prime}  f_j^{ f_j}
{\rm Tr}  \{ \hat \rho  \sum_i\nolimits^{\prime} |y_i\rangle \langle
y_i| \}
  \le  \prod_j\nolimits^{\prime}  f_j^{ f_j}.
\label{UG}
\end{equation}
This represents a state--independent upper  bound.
The necessary  condition for equality sign in (\ref{true1}) is
 given by   the conditions
$
 \langle y_i | \hat \rho| y_i\rangle/a_i  = {\rm const}
$
for any $i,$
whereas the equality sign appears in relation (\ref{UG})  for complete
measurements. These relations  together with the  normalization of relative
frequencies   tend to the necessary  condition for searched state $\hat
\rho$
\begin{equation}
  \langle y_i|  \hat \rho |y_i \rangle   =   f_i .
  \label{false}
 \end{equation}
This is nothing else as the experimental counterpart of the  relation
(\ref{mereni}) and hence  the  starting point of reconstruction based on
inversion. The relation (\ref{false}) may be simply inverted
in the case of orthogonal measurements, which may be considerd as
complete on the given sub--space, tending to the solution
\begin{equation}
 \hat \rho_{f} = \sum_i^{\prime}\nolimits f_i |y_i \rangle \langle y_i|.
\label{ortog}
\end{equation}
Unfortunately, such measurements do not  reveal information about
full density matrix since the nondiagonal elements are lost, as for example
in the case of  particle number measurement.
Techniques dealing with orthogonal measurements
 are therefore not suitable for full  state reconstruction, which
 should be based on the usage of
  nonorthogonal states. On the other hand, in these cases
the completeness and the  existence of  a solution  of the
equation (\ref{false})
cannot  be guaranteed.
Quantum analogy of Gibbs inequality corresponds
to overestimated upper bound   and  tends to  the
conditions imposed by reconstruction techniques.

{\em Maximum likelihood estimation}

\noindent
The problems with existence of a state  achieving the upper bound
descends  obviously  from the fixing of the auxiliary parameters $\bf a.$
The remedy is to keep them free as a subject of further optimization.
For any positively defined operator $\hat B = \sum _i
\lambda_i |b_i\rangle \langle b_i |$ and density operator $\hat \rho$
the simple Lemma     holds
\begin{equation}
{\rm Tr} (\hat \rho \hat B) \le  \max_i \lambda_i .
\label{lemma}
\end{equation}
 The quality sign  is achieved for density matrix   corresponding to
 the  spectral  projector of operator $\hat B$  with maximal eigenvalue.
Using this Lemma the estimation of the  right hand side of the
inequality (\ref{true1}) than reads
\begin{equation}
(\ref{true1}) \le   \lambda({\bf y}, {\bf a})
\prod_i\nolimits^{\prime}  a_i^{ f_i}.
\label{true3}
\end{equation}
 where    $ \lambda ({\bf y}, {\bf a})  $ denotes  formally the
 maximal  eigenvalue of the operator  $ \hat R({\bf y}, {\bf a})$
 with the corresponding eigenvector
 $| \psi({\bf y}, {\bf a}) \rangle .$
 Equality  signs in the chain of inequalities
 are achieved simultaneously if and only if
 \begin{equation}
 \frac{| \langle y_i|  \psi({\bf y}, {\bf a})\rangle |^2}{ a_i} =
 {\rm const,}
 \label{vysledek1}
\end{equation}
independently on the  index  $i .$ Finally,  maximum likelihood
estimation  determines the desired state as
$| \psi({\bf y}, {\bf a}) \rangle,$ where
vector  ${\bf a}$ is given by the solution of the set of
nonlinear equations  (\ref{vysledek1}).
The uncertainty of such a  quantum state estimation may be, according to
the Bayesian formulation  \cite{Jones},   characterized by the likelihood
functional (\ref{likeli}). Since the interpretation of the  probability
distribution on the space of states is rather complicated, the
uncertainty of the prediction may be involved in an  alternative  way.
 The measured data are fluctuating according to the  distribution
 function $P({\bf y}) $ depending on the true state of the system.
Fluctuations of quantum state estimates     may be
represented by the sum of independent   contributions
\begin{equation}
\hat \rho_{MLE} = \langle  \;   | \psi({\bf y}) \rangle
\langle \psi({\bf y})|   \:  \rangle_{\bf y} = \int d {\bf y} P({\bf
y}) | \psi({\bf y}) \rangle
\langle \psi({\bf y})|.
\label{vysledek2}
\end{equation}
This  density matrix
shows how closely the maximum likelihood method allows to estimate an
unknown state  hidden in the  measured  statistics $P({\bf y})$.
 Unfortunately, the proposed method is rather complicated and
examples of  reconstructions specified
above should be solved  separately case by case.
Considerable technical  difficulties may be
 caused, for example,   by possible degeneracy of
operator  $\hat B$   reflecting the structure of performed quantum
measurement.
This particular questions are beyond the scope of this contribution
and represent an  advanced program for further re--interpretation
of existing reconstruction techniques.

Developed technique may be illustrated on
simple but theoretically worth examples.
Quantum state reconstruction  after  the measurement of a Hermitian
operator with orthogonal  spectrum is  the simplest  problem.
Solution corresponds to the application of Gibbs
inequality, since the relation  (\ref{false})  may be solved in this
case. Quantum  state is then reconstructed after each measurement
by the density matrix   (\ref{ortog}).
This is a consequence of the possible degeneracy of the operator
 $\hat B$ mentioned above.
The treatment based on the Gibbs inequality   is overestimated in
general. Provided that (\ref{false})   is   fulfilled in some special
cases, then the solution
should coincide with the prediction of maximum likelihood estimation.

Simple are also the cases of strongly underdetermined data, when
the  state is estimated after single detection $n=1$.
Assume for concretness the standard  ``measurement of Q--function"
  corresponding to
detection of  coherent states $  |y\rangle  =
e^{y \hat a^{\dagger} - y^* \hat a} |0\rangle.$
If  the value $y$ is detected, the  system is
 with the highest likelihood just in the
 state $|y\rangle. $
Provided that system {\em was in coherent state}  $|\alpha \rangle,$
 the output fluctuates as  $ |\langle \alpha|y\rangle |^2/\pi. $
Estimation after  single detection then
 yields the density matrix  of superposition of coherent
 signal  $\alpha$
 and the thermal noise \cite{Perina} with mean number of particles equal to $1,$
$$
\hat \rho_{MLE} =  \frac{1}{\pi} \int d^2 y  e^{-|y-\alpha|^2}
|y\rangle \langle y|.
$$
Difference between the true state  and its estimation is
negligible in the case of classical fields,
but considerable in quantum domain.

Estimating the quantum state after multiple
detection of coherent states, the matrix $\hat R$ should be
diagonalized.
Using the assumption for eigenstates  as
$ |\varphi \rangle  = \sum\nolimits_i^{\prime} V_i
|y_i\rangle, $  linear equations
for desired coefficients $V_i$ and eigenvalues $\lambda $  follow as
\begin{equation}
\frac{f_k}{a_k}  \sum\nolimits_i^{\prime} V_i C_{ki} = \lambda V_k,
\end{equation}
where $C_{ki}=C^*_{ik} = \langle y_k | y_i \rangle, C_{ii}=1.$
This solution determines the coefficients   $\bf a$ according to
the relation  (\ref{vysledek1})  as  $
|\sum\nolimits_i^{\prime} V_i C_{ki}|^2/ a_k ={\rm const} $ for any
index $k.$
Let us illustrate this strategy on the case of double detection
 $n=2$ yielding the values $y_1$ and $ y_2.$ Parameters are
given as $f_1= f_2= 1/2$ and
without loss of generality
$ a_1= 1, a_1/a_2 =x.$
The secular equation for  maximal eigenvalue $\lambda$ reads
$
\lambda^2 - (1+x) \lambda + x- x|C_{12}|^2  =0,
$
yielding easily  solutions for maximal eigenvalue
and its eigenvector.
Equations (\ref{vysledek1}) impose single condition as
$
|C_{12}| \lambda  = \sqrt{x} ( \lambda - 1 + |C_{12}|^2).
$
This nonlinear  system of equations
may be easily solved yielding  expected  solution as $\lambda = 1 + |C_{12}|,
x = 1.$ Projector is given by the normalized Schr\"{o}dinger
 cat--like  state
$$|\varphi\rangle = \frac{1}{\sqrt{2(1 + |C_{12}|)}}
\bigl(e^{i\arg C_{12}}|y_1\rangle + |y_2\rangle   \bigr).$$
Density matrix  ``reconstructing" the coherent state is then
given as
$$\hat \rho_{MLE} =\frac{1}{\pi^2} \int d^2 y_1 d^2 y_2
e^{-|y_1-\alpha|^2-|y_2-\alpha|^2}
|\varphi\rangle \langle\varphi|.
$$
Proposed method  describes  easily  the cases, where the data seem to be
 underdetermined. There is
 also a strong effort to apply  the developed technique to
 the case of large data sets
 estimating   properly the quantum state in the cases of
realistic measurements.

Even if  the problem of positive definiteness used for motivation
may seem   as nit--picking, it has  far-reaching consequeces.
Reconstruction methods based on inversion
of detected data  are  valid only if complete
 information is available and fail, if the information is
limited by  quantum theory.
The method based on maximum likelihood suggests, how to treat the state
reconstruction in this quantum domain.
Since  quantum state comprises  maximum possible  information
about the system, its   proper description is of  fundamental interest.
For example, the strategy of indirect measurement observing
primarily the wave function  and   deriving   all
information about desired variable consequently, represents a general scheme
for an  universal ``measurement of everything".
The fee paid for such an observation should be obviously  the
 accuracy of the detection of a particular observable, since the
 observation of wave function involves the registration of
 non--commuting variable, too.
 The fundamental distinction between universal and accurate
 measurements  may easily disappear provided that an improper
 description of quantum objects is used. Particularly, this is just the
 case of existing reconstruction techniques,
 where  ``negative probabilities"
 may appear as a consequence of inadequate semiclassical treatment.

I am grateful to  Prof. H. Rauch  for the hospitality of Atominstitut
der \"Osterreichischen Universit\"aten and to T. Opatrn\'y for
discusions concerning state reconstruction. This work was supported by the
East--West program of  the Austrian Academy of Sciences and
by the grant of Czech Ministry of Education VS 96028.


\begin{thebibliography}{999999}
\bibitem{note}
 Positive definity of $\hat \rho, $
 i.e. the property $ \langle x |\hat \rho |x\rangle  \ge 0 $ for
 any state $|x \rangle $,   is  very complicated  for testing.
 It  has not to be confused
with the positivity of the Wigner function, which need not be positive!
\end{thebibliography}
\end{document}